\newcommand{\version}{arXiv:\ \ v4\ (final),\ \ \today}

\documentclass[12pt,
onecolumn,%
oneside,%
floats,%
aps,%
prd,%
nobibnotes,%
nofootinbib,%
amsmath,%
amssymb,%
amsfonts,%
superscriptaddress,%
]{revtex4}

\usepackage[utf8]{inputenc}
\usepackage{graphicx,array,dcolumn}
\usepackage{cases}
\usepackage[paperwidth=210mm,paperheight=297mm,centering,hmargin=1.8cm,vmargin=2.5cm]{geometry}
\usepackage{enumerate}

\usepackage{hyperref}

\usepackage[normalem]{ulem}
\usepackage{soul}
\usepackage{bm}
\usepackage{bold-extra}

\def\fun#1#2{\lower3.6pt\vbox{\baselineskip0pt\lineskip.9pt
\ialign{$\mathsurround=0pt#1\hfil ##\hfil$\crcr#2\crcr\sim\crcr}}}

\newcommand{{\SD}}{\rm SD}

\newcommand{{\Mc}}{\mathcal{M}}

\newcommand{\ver}{\mbox{\boldmath${\rm r}$}}

\newcommand{\vep}{\mbox{\boldmath${\rm p}$}}
\newcommand{\veq}{\mbox{\boldmath${\rm q}$}}

 \newcommand{\veA}{\mbox{\boldmath${\rm A}$}}

\newcommand{\vek}{\mbox{\boldmath${\rm k}$}}

\newcommand{\vej}{\mbox{\boldmath${\rm j}$}}

\newcommand{\veB}{\mbox{\boldmath${\rm B}$}}

\newcommand{\veE}{\mbox{\boldmath${\rm E}$}}

\newcommand{\lan}{\langle}
\newcommand{\ran}{\rangle}

\def\-g{\sqrt{-g}}
\newcommand{\venab}{\mbox{\boldmath${\rm \nabla}$}}
\newcommand{\be}{\begin{equation}}
\newcommand{\ee}{\end{equation}}

\newcommand{\ben}{\begin{equation*}}
\newcommand{\een}{\end{equation*}}

\newcommand{\bea}{\begin{eqnarray}}
\newcommand{\eea}{\end{eqnarray}}

\renewcommand\rho{\varrho}
\renewcommand\tilde{\widetilde}

\begin{document}
 
\vspace*{2cm}

\title{\sc \Large{Pre-critical soft photons emission from quark matter}}

\thanks{\version}

\author{\firstname{B.O.}~\surname{Kerbikov}\medskip}

\email{borisk@itep.ru}

\affiliation{ NRC ``Kurchatov Institute'' -- ITEP,  Moscow 117218, Russia \smallskip}

\affiliation{Lebedev Physical Institute, Moscow 119991, Russia \smallskip}

\affiliation{Moscow Institute of Physics and Technology, Dolgoprudny 141700,
Moscow Region, Russia \bigskip}

\date{\today}

\begin{abstract}
\noindent We compute the soft real photon emission rate from the QCD matter in the  vicinity of the critical line at moderate  density and the  temperature approaching the critical one from above. The  obtained production rate exhibits a steep rise  close to $T_c$ due to the formation of the slow fluctuation mode.

\end{abstract}

\maketitle

\large
\section{Introduction}

Heavy ion collision experiments carried out at RHIC and LHC over the last two decades brought about the discovery of a new form of matter with unexpected properties. Several probes are used to  reveal its nature and characteristics. A special role  is played by direct photons. They are produced at all stages of the fireball evolution and  can easily escape the collision region without  reinteracting. Photons and  dileptons production has been studied both experimentally and theoretically for quite a long time. The basic theory concepts have their roots in the studies performed several decades ago \cite{1,2,3,4,5}. The current   status of the field is presented in the  review article \cite{6}.  In this work we consider the real soft photon emission rate from dense quark matter with the temperature approaching the critical one from above.
Real photons means that $q^2=\omega^2-\veq^2=M^2=0,$ and soft corresponds to $\omega \ll T$. For the dilepton production $M^2>0$ is the invariant mass of the lepton pair. This process is not considered in the present work.
Necessary to emphasize that only the external photon is assumed to be soft but the internal momenta in the self-energy diagram may be hard. In a sense the picture is reminiscent of the hard thermal loop approximation. The role of high $T$ is played by the high chemical potential. The dominant contribution to the photon polarization operator comes from the vicinity of the Fermi surface. Up to now the soft photon emission has been  predominantly studied for hot and low density QGP. In this  region of  the QCD phase  diagram perturbative methods including the hard  thermal loop are the  adequate  research  tools \cite{7,8,9,10,11}. Results of several lattice calculations at zero chemical potential are also available \cite{12,13,14}. On the other hand during the last years it became clear that  except for high temperature and low   density  domain  the quark matter is a strongly coupled medium \cite{15}. There are very few calculations of the photon production beyond, or partly beyond, the  perturbation theory \cite{16,17,18,19}.

The reason is that the  finite temperature  retarded self-energy of  virtual photon is known only in perturbation theory \cite{20,21}. Probably the most intriguing  region of the phase diagram lies in the vicinity of the critical temperature at nonzero density. The corresponding  research  program is planned at NICA and FAIR. In this 
domain the  correlation functions are characterized by the  presence of a soft mode of the  fluctuation field. 

The importance of the collective mode in the precritical region of the quark matter at finite density and its relevance for the dilepton production was to our knowledge first pointed out in \cite{22,23}

It will be shown below that the propagator of the fluctuation mode (FP) has the form \be  L(\veq, \omega) =\frac{N}{\frac{T-T_c}{T_c} - i \beta \omega + \xi^2 \veq^2}. \label{1}\ee
The  quantities $N$, $\beta$ and $\xi^2$ will be determined in what follows. One may recognize in (\ref{1}) the linear  response  function of the phase transition theory \cite{24,25}. At small $\omega$ and $\veq^2$ and close to $T_c$ the FP (\ref{1}) can be  arbitrary large and is rapidly varying due to the $(T-T_c)/T_c$ term. We shall evaluate the soft photon emission rate close to $T_c$ using the  expression for the retarded self-energy containing two FP-s. This will lead to the enchanced soft photon production rate. 

The  organization of this paper is as follows. In Section II, we show that there is a rather wide fluctuation region above the critical line at moderate density. In Section III, using the time-dependent  Ginzburg-Landau  functional with Langevin forces we  derive the propagator of the soft collective mode. In Section IV, we address the retarded photon self-energy in the fluctuation region. In Section V, we compute the soft photon emissivity and confront it with the  electrical conductivity computation. We summarize and conclude in Section VI.

\section{Critical fluctuations}

Our focus  in this work is on the finite density pre-critical fluctuation region with $T\to T_c$ from above. Comprehensive study has shown that at high density and low temperature the  ground state of QCD is color superconductor  \cite{26,27}. We consider the 2SC color superconducting phase when $u$- and $d$- quarks participate in color antitriplet pairing but the density is not high enough to involve the heavier $s$- quark. The value of the quark chemical potential under consideration is $\mu\simeq 300$-$400$ MeV and the critical temperature $T_c \simeq 40$-$50$ MeV. The corresponding density is two or three times the normal nuclear density. Both numbers should be considered as educated guess since they rely on model calculations. Similar choice of parameters has been adopted in \cite{22}, namely $\mu \simeq 400$-$500$ MeV and $T_c \simeq 40$-$60$ MeV. Prior to forming a condensate the system goes through the phase of the pre-formed fluctuation quark pairs. In its basic features this state is very different from the fluctuation regime of the BCS superconductor \cite{28}. In the BCS the border between the normal and the superconducting phases is very sharp. In color superconductor it is significantly smeared. Two interrelated explanations of this difference may be given. First, in the BCS the characteristic pair correlation length $\xi$ is large, $\xi \simeq 10^{-4}$ cm, so that $n^{1/3}_e\xi \gg 1$, where $n_e \sim 10^{22}$ cm$^{-3}$ is the electron density \cite{29}. The pairs strongly overlap. In color superconductor the pairs which form the condensate are much more compact and have a small overlap (the Schafroth Pairs, \cite{30}). The role of the  correlation length $\xi$ is taken by the root-mean-square radius $\rho \sim 1$ fm of the quark pair. The 2SC quark matter density $n_q$ is (2-3) times  the normal nuclear density, so that $n_q^{1/3} \rho \sim 1$. Note that $n^{1/3} \xi$ is the BCS-BEC crossover parameter \cite{31,32,33,34,35,36}. Therefore one may say  that at $\mu\sim 300$-$400$ MeV, $T\sim 40$-$50$ MeV the  system is in the crossover regime \cite{28}. The second way to reveal the  difference between the BCS and color superconductor is to  compare the relative values of the energy parameters in the two theories. In the BCS the following scales hierarchy holds $\Delta: \omega_D: \varepsilon_F\simeq 1:10^2:10^4$, where $\Delta\sim  T_c \sim 10^{-4}$ eV is the gap/critical temperature, $\omega_D\sim 10^{-2}$ eV is the Debye energy, $\varepsilon_F \sim 2$ eV is the Fermi energy \cite{29}. In color superconductor the relation is very different,  $\Delta: \Lambda:\mu\simeq 1:8:4$, where $\Delta \sim 0.1$ GeV is the gap , $\Lambda \sim 0.8$ GeV is the UV cutoff, $\mu \sim 0.4$ GeV is the quark chemical potential \cite{35}. The width of the fluctuation region and the  fluctuation contribution to the physical quantities are characterized by the Ginzburg-Levanyuk parameter \cite{24,28,29,35,37,38,39}. There are several definitions of this quantity  in the literature \cite{24,29,37}. The underlying requirement is that the fluctuation corrections to the physical quantities (e.g., the heat capacity, the electrical conductivity) must be much smaller than the characteristic values of these quantities. According to the original Ginzburg estimate \cite{39} based on the fluctuation heat capacity of the BCS superconductor the temperature interval within the fluctuation contribution is essential is 
\be \operatorname{Gi} \simeq \frac{\delta T}{T_c} \sim \left( \dfrac{T_c}{E_F} 
\right)^4, \label{2}
\ee
where $E_F$ is the Fermi energy. To adjust this estimate to the quark matter we replace $E_F$ by $\mu$, use the BCS theory estimate $\xi \sim T_c^{-1}$ \cite{29,37} and then replace $\xi$ by the quark pair radius $\rho$. It should be noted that the rigorous calculation of the pair size in the nonperturbative QCD region is hardly possible. The energy spread of the correlated pair of quarks is $\delta E \sim \Delta \sim 100$ MeV, quarks are relativistic, hence $\delta p \sim \Delta$, and therefore $\rho \sim 1/\Delta \sim 2$ fm. Using the Klein-Gordon equation for the quark pair \cite{40} one can obtain an estimate $\rho \simeq (\sqrt3 \Delta)^{-1} \sim 1$ fm. Equation (\ref{2}) describes the universal dependence of $\operatorname{Gi}$ on the superconductor physical parameters. Depending on the specific properties of a given material it should be supplemented  by an additional numerical factor \cite{24,37}. The evaluation of this factor for the quark matter is a difficult problem. We shall not try to solve it since the equation (\ref{2}) contains a strong fourth power dependence on $T_c$, $\mu$, $\rho$, and the overall numerical coefficient is less important. As we discussed above the values of these parameters are not narrowly limited. Replacing in (\ref{2}) $E_F$ by $\mu$ and using the estimate $\rho \sim T_c^{-1}$ we write the following two complementary expressions for the Ginzburg parameter
\be \operatorname{Gi} \simeq \frac{\delta T}{T_c} \simeq \left( \dfrac{T_c}{\mu} 
\right)^4 \simeq (\mu\rho)^{-4}, \label{3}
\ee
Due to the fourth power dependence on $T_c$, $\mu$ and $\rho$ and due to some uncertainty in their values we can estimate only the reliable interval of the $\operatorname{Gi}$ parameter. For $T_c\simeq(40$-$50)$ MeV, $\mu \simeq (300$-$400)$ MeV, $\rho \simeq (1$-$2)$ fm the quantity $\operatorname{Gi}$ varies from $10^{-4}$ to $10^{-2}$. We remind that for the ordinary superconductors $\operatorname{Gi} \sim 10^{-14}$-$10^{-12}$ \cite{29,37}. In the next Section we shall discuss the bound on $\operatorname{Gi}$ from below. 

\section{Collective mode propagator} 

The FP of the form (\ref{1}) may be derived in several ways. In \cite{41} it was obtained by solving the Dyson equation with relativistic Matsubara quark propagators. Here we shall use the time-dependent Ginzburg-Landau (GL) functional \cite{42,43} with the stochastic Langevin forces. The approximations and omissions in the derivation to follow will be discussed at the end of this Section. In absence of the external  electromagnetic field the time-dependent GL equation for the  fluctuating pair field $\Psi(\ver,t)$ reads

\be -\gamma \frac{\partial}{\partial t} \Psi (\ver , t) = \frac{\delta F [\Psi]}{\delta \Psi^*} + \eta (\ver , t).\label{4}\ee

here $\gamma$ is the order parameter relaxation constant, $\eta (\ver, t)$ are the Langevin forces. The GL functional with the quartic term dropped (see below) has the form \cite{24,29,37}
\be F[\Psi] =\nu \int [ \varepsilon |\Psi (\ver, t)|^2 + \xi^2 |\venab \Psi (\ver,t)|^2] dVdt, \label{5}\ee
where $\nu =\mu p_F/\pi^2$ is  the relativistic  density of  states at the Fermi  surface \cite{28,41}, $\varepsilon = (T-T_c)/T_c$, $\xi$  is the coherence length which may be expressed in terms of the  diffusion  coefficient as $\xi^2 = \frac{\pi}{8T} D$ \cite{37,41}. Addressing the readers to the above references we present a sketch of the derivation. The starting point is the QCD partition function. Expanding it in powers of $|\Psi|^2$ one arrives at the needed GL expression. The term $\xi^2 |\venab \Psi|^2$ in (\ref{5}) enters into this expression with the coefficient equal to \cite{28}
\be
\xi^2 = \dfrac{7\zeta(3)v_F^2}{48\pi^2 T^2}\chi\left(\dfrac{1}{2 \pi T \tau}\right)
\label{6}
\ee
Here $\tau$ is the momentum relaxation time. The function $\chi(z)$ is
\be
\chi(z)=\dfrac{8}{7\zeta(3)}\sum\limits^{\infty}_{n=0} \dfrac{1}{(2n+1)^2(2n+1+z)} \longrightarrow 
\begin{cases}
1,&\text{$z \to 0$;}\\
\frac{\pi^2}{7\zeta(3)}z^{-1},&\text{$z \gg 1$}.
\end{cases}
\label{7}
\ee

The relaxation time $\tau$ depends on the temperature, density and the quark flavor. It can be also identified with the mean free path time, or the relaxation time in the Boltzmann approximation. The reliable estimation of $\tau$ is absent even for $\mu = 0$. For example, in \cite{44} it varies at $\mu = 0$ in the interval $\tau \simeq (0.1\text{-}0.9)$ fm. Therefore, let us consider the two limining  cases, namely  $2\pi T\tau \ll 1$ and $2\pi T\tau \gg 1$. For the critical temperature under consideration $T_c \simeq (40\text{-}50)$ MeV the two limits take place at $\tau \lesssim 0.3$ fm and $\tau \gtrsim 2$ fm correspondingly. Based on our experience in the calculation of the quark matter conductivity \cite{41} we consider the choice $\tau \lesssim 0.3$ fm more realistic. In the above two limits one obtains correspondingly 

\be
\xi^2 \simeq \dfrac{\pi}{8T}\left(\frac{1}{3}v_F^2\tau\right) \equiv\dfrac{\pi}{8T}D_1,
\label{8}
\ee
\be
\xi^2 \simeq \dfrac{\pi}{8T}\left(\frac{v_F^2}{6\pi T}\right) \equiv\dfrac{\pi}{8T}D_2.
\label{9}
\ee
The quantity $D_1$ is a standard diffusion coefficient $D_1 \sim vl$.
The coefficient $D_2$ has a meaning of a diffusion coefficient in the quasi-free ballistic regime \cite{37}. It can be obtained from (\ref{8}) by the replacement $\tau \to (2 \pi T)^{-1}$. We consider a rather dense quark matter. It is in a collisional ``dirty'' regime, not in a ballistic one. Therefore in our calculations we shall take $\xi^2$ in the form (\ref{8}), omit the lower subscript, and slightly vary the parameter $\tau$. 

Now we perform a Fourier transform to momentum space

\be \Psi(\ver, t) = \int \frac{d\veq}{(2\pi)^3} \frac{d\omega}{2\pi} e^{i\veq\ver-i\omega t}\phi(\veq,\omega).
\label{10}\ee

The GL functional in momentum space reads
\be
F[\phi] = \nu \int \frac{d\veq}{(2\pi)^3} \frac{d\omega}{2\pi} \left[ \left( \varepsilon + \frac{\pi}{8T}D q^2 \right) |\phi(\veq, \omega)|^2 \right].
\label{11}\ee

The time-dependent Eq.(\ref{4}) takes the following form in momentum space
\be 
-\left[ - i \gamma \omega + \nu\left(\varepsilon + \frac{\pi}{8T}D\veq^2\right) \right] \phi (\veq, \omega) =\eta (\veq, \omega).
\label{12}\ee

The solution of (\ref{12}) may be written as 
\be \phi(\veq, \omega) = L (\veq , \omega) \eta (\veq , \omega), \label{13}\ee
where \be L (\veq , \omega)=-(-i\gamma \omega + \Omega_{\veq})^{-1}\label{14} 
\ee
with $\Omega_{\veq} = \nu \left( \varepsilon + \frac{\pi}{8T} D q^2 \right)$. To ascertain that $L$ is actually the fluctuation mode propagator we must verify that it satisfies the fluctuation - dissipation theorem \cite{24,45}. The theorem states that the equal time correlator $\lan \Psi (\ver , t) \Psi^*(\ver', t)\ran$ is expressed via the retarded propagator. 
The solution (\ref{13}) satisfies this requirement provided the correlator of the Langevin forces have a gaussian white noise form in the coordinate space
\be 
\lan \eta (\ver, t) \eta^* (\ver', t') \ran = 2 T \gamma \delta (\ver -\ver') \delta (t-t').\label{15}\ee
Then
\be \lan \eta (\ver, t) \eta^* (\ver', t) \ran = 2 T \gamma \int \frac{d\veq}{(2\pi)^3} e^{-i \veq (\ver-\ver')}\int  \frac{d\omega}{2\pi},
\label{16}\ee
and
\be
\lan \Psi (\ver, t) \Psi^* (\ver', t) \ran = 2T\gamma \int \frac{d\veq}{(2\pi)^3} e^{i\veq (\ver-\ver')}\int^\infty_{-\infty} \frac{d\omega}{2\pi} L(\veq, \omega) L^*(\veq, \omega).
\label{17}\ee
Therefore, $\lan \Psi (\ver, t) \, \Psi^* (\ver', t) \ran$ in momentum space is 
$$ \lan \Psi (\ver, t) \, \Psi^* (\ver', t) \ran_{\vep} = 2T\gamma \int^\infty_{-\infty}  \frac{d\omega}{2\pi} L (\veq, \omega) L^* (\veq, \omega) =$$
\be  
=- \int \frac{d\omega}{2\pi} \frac{2T}{\omega} \operatorname{Im} L(\veq, \omega). 
\label{18}\ee
Thus, $L(\veq, \omega)$ given by (\ref{14}) meets the needed requirement. The last step is to express the coefficient $\gamma$ in terms of other parameters. From (\ref{12}) it follows that the relaxation time of fluctuations with momentum $\veq^2$ is  
\be \tau = \dfrac{\gamma}{\nu \left( \varepsilon + \frac{\pi}{8T}D\veq^2 \right)}.
\label{19}
\ee
Keeping in the  denominator of (\ref{19}) only the term $\nu\varepsilon$, which is equivalent to retaining only the term $\varepsilon |\psi|^2$ in (\ref{5}), and comparing the result with the GL decay time $\tau_{GL} = \pi [8 (T-T_c)]^{-1}$ \cite{29,37} we obtain $\gamma = \pi\nu/8T_c$. This completes the  derivation of the FP
\be L(\veq, \omega)=-\dfrac{1}{\nu\left[\varepsilon + \frac{\pi}{8T} (-i\omega + D \veq^2)\right]}.\label{20}\ee
We note that since we consider the temperature interval $\varepsilon \lesssim 10^{-2}$, the temperature $T$ in (\ref{20}) may be replaced by $T_c$ without a noticeable loss of accuracy. We do not find this simplification necessary. The way the above results were obtained may rise questions about the validity of the employed approximations.

Let us discuss the debatable points. The electromagnetic field  was not included into Eq.(5). It is well known that the external magnetic field applied to the superconductor gives rise to important phenomena like Meissner effect. It also influences the physics of fluctuations in ordinary superconductors \cite{37} as well as in color superconductors \cite{28}. Quark matter may be embedded into magnetic field when it is produced in the peripheral collisions of the ultra-relativistic heavy ions collisions at RHIC and LHC \cite{46}. Quark-gluon matter formed in such collisions has high temperature and low density which excludes the formation of the color quark confinement. The present investigation may be important for the future experiments at NICA and FAIR where the sizable magnetic field, if any will not be generated. The omission of the fourth order term in (\ref{5}) is a subtle question. Without this term (\ref{5}) corresponds to the Gaussian fluctuations with no interaction between them. In the immediate vicinity of $T_c$ at $\varepsilon \lesssim \operatorname{Gi}$ this approximation breaks down \cite{24,29,37}. Here one encounters a difficult problem. Renormalization group method is used in this critical region \cite{24,37}. However in the three dimensional case the complete solution is lacking and we shall not dwell on that. Based on the values of $\operatorname{Gi}$ obtained in the previous Section we shall present the results down to $\varepsilon \simeq 10^{-4}$ keeping in mind that below $\varepsilon \simeq 10^{-2}$ the corrections due to the interaction between fluctuations may come into play. 

\section{The photon pre-critical self-energy}

To calculate the photon emission rate we have to construct the photon self-energy operator in the pre-critical region. Intensive studies since $1960$-s resulted in a fairly complete picture of the fluctuation effects near $T_c$ -- see \cite{37} and a long list of references therein. Three basic papers \cite{47,48,49} should be singled out of this list. Worth mentioning also Ref.\cite{50} in which the fluctuation conductivity has been studied in the strong coupling limit. The quark pairs under study in this work are in the strong coupling regime close to the BCS-BEC crossover \cite{28,36}.

According to the diagram calculus the self-energy $\Pi(\veq,\omega)$ in the pre-critical region can be constructed from the two kinds of the building blocks. These are the quark Matsubara Green's functions $G(\vep,\varepsilon_n)$ (see below) and the fluctuating field pair average represented by the FP (\ref{20}). The GL funtional (\ref{5}) without the fourth order term describes an almost free field. For the free field the Wick's theorem states that the higher order correlators are expressed as products of the pair averages, i.e. the FP-s. Therefore we are really left with the two above building blocks. Attributing the solid lines to the quark propagators and the wavy lines to the FP-s we come to the set of diagrams for the photon self-energy (the retarted Green's function). The possible diagrams have been discussed in a vast number of works, see \cite{37} for the review and \cite{47,48,49} for the original results. The two diagrams which were compared in a number of publications are  the Aslamazov-Larkin \cite{47} and the Maki-Thompson \cite{48,49} ones. It is beyond the scope of this paper to reproduce their comparative analysis \cite{29,37,51}. The bottomline is that the theoretical arguments supported  by the experimental data \cite{52} allow to conclude that the dominant role is played by the celebrated  Aslamazov-Larkin (AL) diagram \cite{37,47,51} shown in Fig.1. 

 
\begin{figure}[!ht]
\centering
\includegraphics[width=0.40\textwidth]{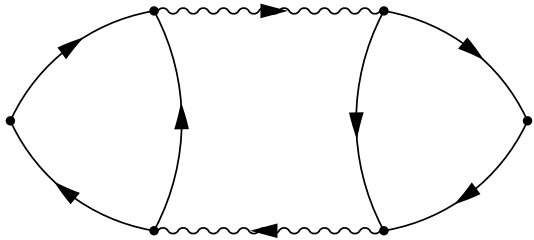}
\caption{The AL diagram for the polarization operator. The solid lines correspond to the Matsubara quark Green's functions, the wavy lines to the FP-s.}
\label{fig01}
\end{figure}


It consists of two quark loops connected by FP-s and reads 
{\small 
\be \Pi_{lm} (\omega=i\omega_k=|\vek|) =-3Q^2 T\sum_{\Omega j} \int \frac{d\veq}{(2\pi)^3} B_l (\vek, \veq, \Omega_j, \omega_k) L(\vek+\veq, \Omega_j + \omega_k) B_m (\vek, \veq, \Omega_j, \omega_k) L( \veq, \Omega_j ).
\label{21}
\ee 
} 
Here $\omega_k$ and $\Omega_j$ are the Matsubara frequencies. The factor 3 comes from color, $Q^2 =\frac59 e^2$ for two flavors, $e^2=4\pi \alpha$.  The trace over the Dirac  indices is included into the 3-vector $\veB$ with components $B_l$ and $B_m$. The factor $\veB$ corresponds to the three Green's functions block. 
Two points concerning Eq.(21) deserve an explanation. The first one is that the dependence of the quark loop $\veB$ on $\vek$ can be dropped out and therefore the self-energy $\Pi$ is a function of $\omega=|\vek|$ as it should be for real photons. Second, the three-vector $\veB$ is by symmetry arguments proportional to $\veq$, $\veB \sim \veq$. The quark loop $\veB$ is given by the following expression
$$ \veB (\vek, \veq, \Omega_j, \omega_k) = T \sum_{\varepsilon_n} \lambda (\veq, \varepsilon_n + \omega_k ,  \Omega_j -\varepsilon_n) \lambda (\veq, \varepsilon_n, \Omega_j - \varepsilon_n)\cdot$$
   \be \cdot \int \frac{d\vep}{(2 \pi)^3} tr_D [ \vec{\gamma} G(\vep ,\tilde\varepsilon_n) G(\vep+\vek, \tilde \varepsilon_n +\omega_k) G ( \veq -\vep, \Omega_j -\tilde\varepsilon_n)].\label{22}\ee 
   
   The Matsubara propagators in (\ref{21}) have the form 
   \be G(\vep, \tilde \varepsilon_n) = \frac{1}{\gamma_0 (i\tilde \varepsilon_n +\mu) - \vec{\gamma}\vep-m}.\label{23}\ee
   where $\tilde \varepsilon_n = \varepsilon_n + \frac{1}{2\tau} \operatorname{sgn} \varepsilon_n$,  $\varepsilon_n =\pi T (2n+1)$, where $\tau$ is the  momentum relaxation time. 
This quantity was already introduced in Sec.III. Alternatively $\tau$ may be called the mean free path time. It enters into the Drude formula for the quark matter conductivity and into the Boltzmann equation in the relaxation time approximation \cite{41}. From the formal point of view,  $\tau$ regulates the pinch (collinear) singulatities. The factors  $\lambda$-s are the vertex renormalization corrections \cite{37,51}.  At $\veq \to 0$, $\omega_k\to 0$ the product of the two $\lambda$-s takes the limiting value  $|2\tilde{\varepsilon}_n|^2/|\varepsilon_n|^2$ \cite{41,51}. The quark loop (\ref{22}) is calculated under the following conditions: (i) $|\vek|<<|\vep|$, and (ii) $T<<\mu$. The first condition is easily recognized as the hard thermal loops approximation HTL. The external momentum $\vek$ is assumed to be soft since we are interested in the soft photon emission, but internal momentum $\vep$ is hard. However, in our case this is not due to the high temperature as in the standard HTL, but due to the fact that the dominant contribution to the above integral comes from the vicinity of the Fermi surface with $\vep \sim \mu$, and $\mu \simeq 300$-$400$ MeV, i.e., high. Therefore we replace in (\ref{22}) $G(\vep+\vek,\tilde{\varepsilon}_n+\omega_k) \simeq G(\vep, \tilde{\varepsilon}_n+\omega_k)$ and $\veB$ becomes $\vek$ independent. By symmetry arguments $\veB \sim \veq$. Integration in (\ref{22}) is  performed using the Fermi surface integration measure
   \be \int \frac{d\vep}{(2\pi)^3} = \frac{\nu}{2} \int \frac{d \Omega_{\vep}}{4\pi} \int^\infty_{-\infty} dt, \label{24}\ee
   where $t=\sqrt{\vep^2+m^2} -\mu$, $\nu=\frac{\mu p_F}{\pi^2}$. In the vicinity of $T_c$ the FP (\ref{20}) has a pole structure due to the $\varepsilon$ term. The dependence of $L(\veq, \Omega_j)$ and $L(\veq, \Omega_j-\omega_k)$ on $\Omega_j$ and $\omega_k$ is much stronger than the dependence of the  Green's functions on the same quantities. We shall keep in the  propagators entering into $\veB (\veq, \Omega_j, \omega_k)$ only the dependence on the fermionic frequencies $\tilde\varepsilon_n$ and  evaluate $\veB (\veq, \Omega_j = \omega_k =0)$. Expanding $G(\veq-\vep,-\tilde\varepsilon_n)$ in (\ref{22}) at $\veq \to 0$, one has 
$$G(\veq-\vep,-\tilde\varepsilon_n) \simeq G(-\vep,-\tilde\varepsilon_n) + \veq \dfrac{\partial}{\partial \vep}G(-\vep,-\tilde\varepsilon_n)=$$
\be
\label{25}
G(-\vep,-\tilde\varepsilon_n) + \dfrac{(\veq\vep)}{\mu} \dfrac{\partial}{\partial t}G(-\vep,-\tilde\varepsilon_n).
\ee
Substituting (\ref{25}) in (\ref{22}) one easily observes that the angular integration kills the contribution of the first term of (\ref{25}). The second term yields 
\be \veB (\veq) =- \nu T \sum_{\varepsilon_n} \frac{|2\tilde\varepsilon_n|^2}{|\varepsilon_n|^2} \int \frac{d \Omega_{\vep}}{4\pi}\frac{ (\veq\vep)\vep}{\mu^2}  \int^\infty_{-\infty} \frac{dy}{(y^2+\tilde \varepsilon_n^2)^2}.\label{26}\ee
   
  Performing the integration and using (\ref{7}-\ref{9}) one gets
  \be \veB(\veq) = - \veq \frac{7\zeta(3)}{12} \frac{\nu}{\pi^2 T^2} \frac{\vep^2}{\mu^2}\chi\left(\frac{1}{2\pi T\tau} \right) = - 4 \veq \dfrac{\pi\nu}{8 T}D,  \label{27}\ee


where $D$ are the diffusion coefficients defined by (\ref{8}). From (\ref{21}) and (\ref{27}) we have
 \be \Pi_{lm} (\omega_k) =-12{Q^2}{T}\left(\dfrac{\pi\nu}{8 T}\right)^2{D}^2 \sum\limits_{\Omega_j} \int \frac{d\veq}{(2\pi)^3} q_l q_m  L( \veq, \Omega_j )  L( \veq, \Omega_j+\omega_k ).\label{28}\ee

 
 To evaluate the sum in (\ref{28}), we can use a technique of replacing the summation (\ref{28}) by the contour integration (the si-called ``Eliashberg trick'') \cite{37,53}
 \be T\sum_{\Omega_j} f(\Omega_j) = \frac{1}{4\pi i} \oint dz\coth \frac{z}{2T} f (-iz),\label{29}\ee
where $z=i\Omega_j$. The contour of integration is depicted in the original work \cite{53} and in \cite{37}. In (\ref{28}) the FP-s are defined over the discrete bosonic Matsubara frequencies. We have to perform the analytic continuation of the FP-s. The retarted one $L^R(\veq, -iz)$ is analytic in the upper half-plane $\operatorname{Im} z > 0$, and the advanced one $L^A(\veq, -iz)$ does not have singularities in the lower half-plane. Note that the FP given by (\ref{20}) is the $L^R$ one. The $L^A$ is obtained by replacing in (\ref{20}) $\omega \to -\omega$. Performing the contour integration \cite{37,53} one gets
{\small
\be \Pi(\omega) =-\frac{\pi Q^2 \nu^2 D^2}{32 T^2} \int \frac{d\veq}{(2\pi)^3}\veq^2 \int^\infty_{-\infty}  dz\coth \frac{z}{2T}\left[L^R(\veq, -iz-i\omega)+ L^A(\veq, -iz+i\omega)\right]{\operatorname{Im}}L^R(\veq, -iz).\label{30}\ee } 
Next we expand the integrand in powers of $\omega$ and subtract the zeroth order term. This may be regarded as imposing the Ward identity. The term linear in $\omega$ reads
$$
L^R(\veq, -iz-i\omega) + L^A(\veq, -iz+i\omega)=-\omega\dfrac{d}{dz}\left(L^R(\veq, -iz) - L^A(\veq, -iz)\right)=$$
\be=-2 i \omega \dfrac{d}{dz}{\operatorname{Im}}L^R(\veq, -iz)
\label{31}
\ee

Substituting (\ref{31}) into (\ref{30}) and integrating by parts we obtain
   
\be \Pi(\omega) =-i\omega \frac{\pi Q^2 \nu^2 D^2}{32 T^3} \int \frac{d\veq}{(2\pi)^3}\veq^2 \int^\infty_{-\infty}  dz\frac{[\operatorname{Im}L^R(\veq, -iz )]^2}{\operatorname{sh}^2\frac{z}{4T}}.\label{32}\ee  

Expanding $\operatorname{sh}^2 \dfrac{z}{4T}$ at $z\ll 4T$ and integrating over $dz$ we obtain

\be \Pi(\omega) =-i\omega \frac{\pi^3  Q^2 D^2}{128 T^2} \int \frac{d\veq}{(2\pi)^3}  \frac{\veq^2}{\left(\varepsilon +\frac{\pi}{8T} D \veq^2 \right)^3}=-i\omega \frac{3Q^2}{64}\left(\frac{8T}{\pi D}\right)^{1/2} \varepsilon^{-1/2}. \label{33}\ee

 As expected, the polarization operator is a  singular function at $T\to T_c$  with the $\left( \frac{T}{T-T_c}\right)^{1/2}$ singularity. 
 
 \section{Photon emission rate}
 
The thermal emission rate of soft photons with energy $\omega$ is related to the retarded photon self-energy as \cite{54}
 \be \omega\frac{dR}{d^3k} = - \frac{2}{(2\pi)^3} \operatorname{Im} \Pi (\omega)\frac{1}{e^{\omega/T}-1}.\label{34}\ee
 Here $\Pi(\omega)$ is the transverse projection of $\Pi^\mu_\mu$, the longitudinal projection vanishes at $\vek=0$. Using (\ref{33}) for $\Pi(\omega)$ we obtain
 
   \be \omega\frac{dR}{d^3k} = \frac{3Q^2 T}{2^{8} \pi^3}\left(\dfrac{8T}{\pi D}\right)^{1/2} \varepsilon^{-1/2}. 
   \label{35}\ee
   
   Equation (\ref{34}) is valid to order $e^2$ in  electromagnetic  interaction and  to all orders in strong interaction. Expression (\ref{35}) corresponds to the diagram shown in Fig.1. It describes the emission of soft real photons with $\omega\ll T$ and is applicable within  the  pre-critical region $10^{-4} < \delta T/T_c \ll 1$. As it was explained in Sec. III corrections due to non-linearity of fluctuations may come into play at $\delta T/T_c \simeq 10^{-4}$. In Fig.2 the photon production rate is plotted as a function of $\varepsilon$ for $T_c = 40$ MeV and $T_c = 50$ MeV and $\tau = 0.1$ fm and $\tau = 0.3$ fm. The main feature of the emission rate (\ref{35}) is its steep rise approaching $T_c$ from above. The dependence on $\tau$ is rather weak and on $T_c$ is not very pronounced.

\begin{figure}[ht] 
\setlength{\unitlength}{1.0cm}
\centering
\begin{minipage}[h]{0.485\linewidth}
{\includegraphics[width=0.99\textwidth]{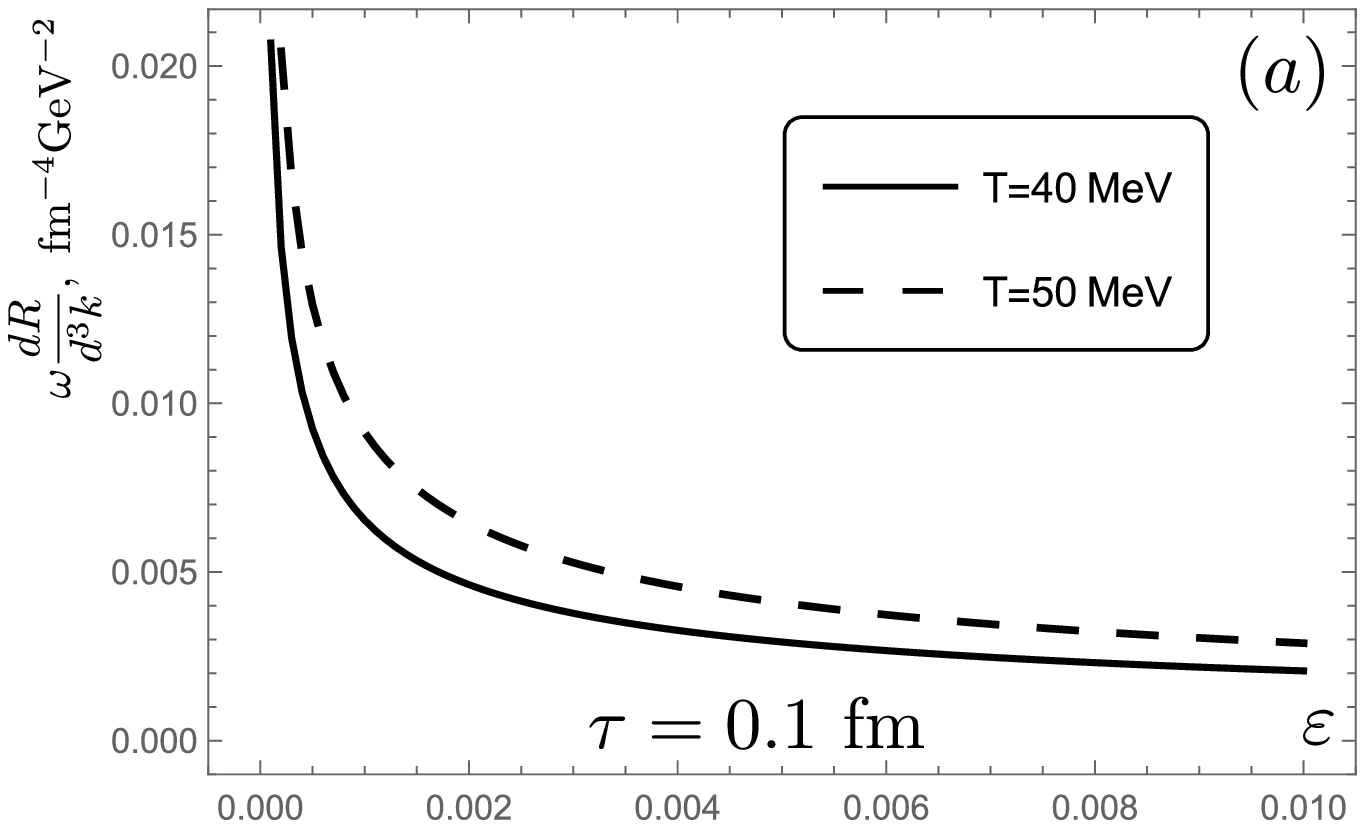}} 
\end{minipage}
\hfill
\begin{minipage}[h]{0.485\linewidth}
{\includegraphics[width=0.99\textwidth]{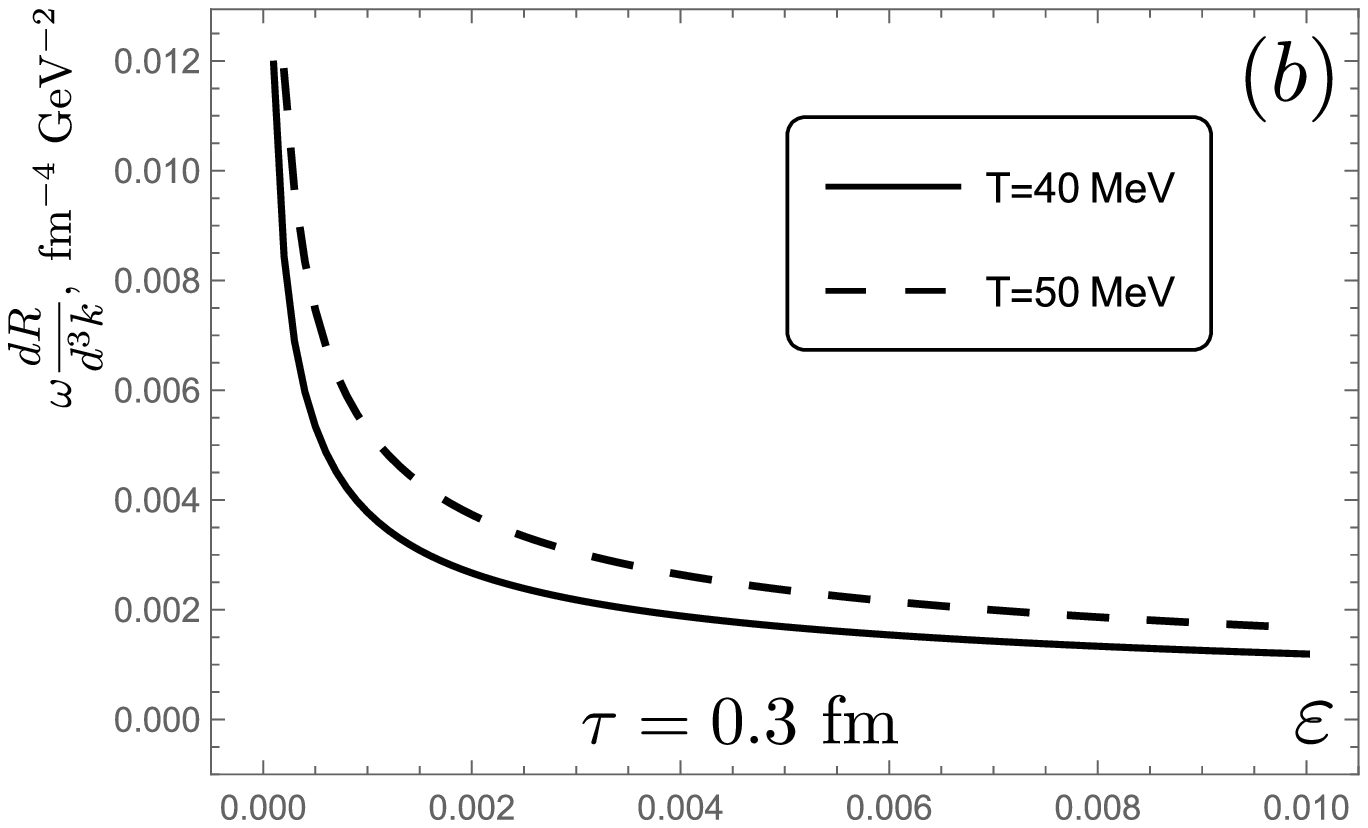}} 
\end{minipage}
\caption{Pre-critical soft photon emission rate. Panel (a): $\tau = 0.1$ fm; Panel (b): $\tau = 0.3$ fm. The solid lines in both panels represent $T_c = 40$ MeV, the dotted lines -- $T_c = 50$ MeV. }
\label{fig02}
\end{figure}

As we mentioned in the Introduction there are very few calculations of the photon emissivity at finite density. There are some common points between our results and that of Ref. \cite{16}. The difference is that in \cite{16} the quark matter is supposed to be in a color superconducting CFL phase with quarks of three flavors $u$, $d$, and $s$ participating in pairing. In this work we consider the precursor virtual pairing of $u$ and $d$ quarks at the temperature just above the critical one for the formation of the condensate. The bird's-eye view is that in \cite{16} the characteristic soft photon emission rate is around $10^{-4}$ fm$^{-4}$ GeV$^{-2}$ (see Fig.12 of \cite{16}) while in our work it is $\sim 10^{-3}$ fm$^{-4}$ GeV$^{-2}$. It means that slow fluctuation mode present in our study enhances the photon emissivity. 

   
The soft photon radiation is closely related to the electrical conductivity of quark matter \cite{12,13,14,55,56}. One can write the following equation for the electric current  \cite{57}
\be
\vej(x) = -\int\Pi(x-y)\veA(y) d^4y.
\label{36}
\ee 
Replacing in Fourier transform of (\ref{36}) $\veA(\vek,\omega)=\veE(\vek,\omega)/i\omega$ and comparing with $\vej=\sigma\veE$ we obtain \cite{37,57}
\be
\sigma(\omega)=-\dfrac{1}{\omega}\operatorname{Im}\Pi(\omega).
\label{37}
\ee 
Comparison of (\ref{35}) and (\ref{37}) gives
\be
\omega\dfrac{dR}{d^3k}=\dfrac{2}{(2\pi)^3}T \sigma(\omega).
\label{38}
\ee 
Note that $\sigma(\omega)$ is of the same order $\alpha$ in electromagnetic interaction as the photon emissivity $\omega\,dR/d^3k$. The appearance of an additional factor $\alpha$ in the right-hand side of (II.16) of \cite{13}, (7) of \cite{55} and (25) of \cite{14} is unclear to the present author. Possibly this is some problem of notations. One finds a large number of the quark matter electrical conductivity calculations in the literature, see, e.g., \cite{41} and references therein. Equations (\ref{35}) and (\ref{38}) yields for $\sigma$ at $T=0.05$ GeV, $\tau=0.2$ fm, $\varepsilon^{-1/2}=20$ the result $\sigma = 0.09$ fm$^{-1}$. This value was previously obtained in our paper \cite{41} dedicated to the electrical conductivity of quark matter. 

\section{Conclusions}

In this paper we have investigated the soft photon emission rate from dense quark matter in the pre-critical region. This part of the QCD phase diagram is up to now to a great extent kept in the dark both from the experimental and theoretical sides. We persued the approach based on the Aslamazov-Larkin diagram which proved to be very successful in condensed matter theory. For quark matter this attitude allowed to describe the transport anomalies near the phase transition temperature \cite{58,59}. In particular, the bulk viscosity diverges near $T_c$ as $\zeta \sim \varepsilon^{-3/2}$ \cite{58}. This is close to the critical behavior $\zeta \sim \varepsilon^{-z\nu + \alpha}$, $z \simeq 3$, $\nu \simeq 0.6$, $\alpha = 0.11$ predicted in $d=4-\varepsilon$ renormalisation, modes coupling, or isomorphism between the quark fluid and 3d Ising system \cite{61,62,63,64}.

The most important feature of the soft photon emissivity rate is its rise when the temperature approaches $T_c$ from above. Close to $T_c$ the fluctuation radiation rate exceeds by an order of magnitude the rate from the color superconducting rate \cite{16}.  The origin of this phenomenon is the formation of the slow fluctuation made in the quark matter. This excitation is described by the fluctuation propagator which is singular at $T_c$ in the limit $\omega \to 0$, $\vek \to 0$. The enhancement of the soft photon production near $T_c$ may be a tentative proposal for the NICA/FAIR investigation.    

\section*{ACKNOWLEDGMENTS}

The author thanks Andrey Varlamov for illuminating discussions and Mikhail Lukashov for useful comments. I am grateful to Prof. Masakiyo Kitazawa and Prof. Teiji Kunihiro for pointing me at Refs.\cite{22,23}. This work was supported by grant from the RFBR number 18-02-40054.

\end{document}